\documentclass[
    ,final            % use final for the camera ready runs
, mathptm
  ]
  {aipproc}

\layoutstyle{8x11single}
\usepackage{amsmath}
\newcommand{\fig}[1]{Fig.~\ref{fig:#1}}

\newcommand{\ord}[1]{\mathcal{O}(#1)}

\newcommand{\df}{\mathrm{d}}

\newcommand{\Tau}{\mathcal{T}}

\newcommand{\cL}{{\mathcal L}}

\newcommand{\bT}{\mathbf{T}}

\newcommand{\vC}{\vec{C}}

\newcommand{\hq}{\hat{q}}
\newcommand{\hs}{\hat{s}}

\newcommand{\hS}{\widehat{S}}

\newcommand{\nn}{\nonumber}

\newcommand{\Ecm}{E_\mathrm{cm}}

\newcommand{\hemi}{\mathrm{hemi}}

\newcommand{\one}{{(1)}}

\newcommand{\id}{\mathbf{1}}

\begin{document}

\title{Jet Regions from Event Shapes and the N-Jet Soft Function at Hadron Colliders}

\classification{13.87.Ce, 12.38.Cy, 13.85.Qk\\
 Talk given by T.T.~Jouttenus at the PANIC 2011
  conference. {\bf Preprint:} MIT-CTP 4290}

\author{Teppo T. Jouttenus}{
  address={Center for Theoretical Physics, Massachusetts Institute of Technology, Cambridge, Massachusetts 02139, USA}
}

\author{ Iain W. Stewart}{
  address={Center for Theoretical Physics, Massachusetts Institute of Technology, Cambridge, Massachusetts 02139, USA},
  altaddress={Center for the Fundamental Laws of Nature, Harvard University, Cambridge, Massachusetts 02138, USA}
  }

\author{Frank J. Tackmann}{
  address={Center for Theoretical Physics, Massachusetts Institute of Technology, Cambridge, Massachusetts 02139, USA}
 }

\author{Wouter~J.~Waalewijn}{
  address={Department of Physics, University of California at San Diego, La Jolla, California 92093, USA}
}

\begin{abstract}
  The $N$-jettiness event shape divides phase space into $N+2$ regions, each
  containing one jet or beam. These jet regions are insensitive to the
  distribution of soft radiation and, with a geometric measure for
  $N$-jettiness, have circular boundaries.  We give a factorization theorem for
  the cross section which is fully differential in the mass of each jet, and
  compute the corresponding soft function at next-to-leading order (NLO).  For
  $N$-jettiness, all ingredients are now available to extend NLO cross sections
  to resummed predictions at next-to-next-to-leading logarithmic order.
\end{abstract}

\maketitle

%%%%%%%%%%%%%%%%%%%%%%%%%%%%%%%%%%%%%%%%%%%%
%% MAINMATTER
%%%%%%%%%%%%%%%%%%%%%%%%%%%%%%%%%%%%%%%%%%%%

% \section{Start}

The measurement of exclusive jet cross sections, where one identifies a certain
number of signal jets but vetoes additional jets, is an important aspect of
Higgs and new-physics searches at the LHC and Tevatron. Since the relative
contributions of various signal and background channels often vary with the
number of hard jets in the event, the sensitivity of the search is improved by
optimizing the analysis for each separate jet bin. Thus, reliable theoretical
calculations of exclusive jet cross sections are essential.

The complication compared to the calculation of an inclusive $N$-jet cross
section, where one sums over additional jets, comes from the fact that the veto
on additional jets imposes a restriction on the energetic initial- and
final-state radiation off the primary hard partons, as well as the overall soft
radiation in the event. This restriction on additional emissions leads to the
appearance of large Sudakov double logarithms in perturbation theory. For
this reason, the calculation of exclusive jet cross sections is traditionally
carried out with parton-shower Monte Carlo programs, where the parton shower
allows one to resum the most singular leading double logarithms.

An alternative analytic approach to calculate exclusive jet cross sections is
possible using factorization and the methods of soft-collinear effective theory
(SCET)~\cite{SCET}.  SCET allows one
to factorize the $N$-jet cross section into pieces depending on only one scale and
resum the large logarithmic contributions. An advantage of this approach is that the resummation can be carried
out to much higher orders than is possible with parton showers.

Schematically, the cross section for $pp \to N$ jets (plus some nonhadronic
final state like a $W$, $Z$, or Higgs if desired) can be factorized as
%%%
\begin{equation} \label{eq:sigmaN}
\sigma_N = H_N \times \Bigl[ B_a B_b \prod_{i = 1}^N J_i \Bigr] \otimes S_N
\,.\end{equation}
%%%
This formula directly applies to observables that implement a veto on additional
jets. The
hard function $H_N$ encodes hard virtual corrections to the underlying partonic
$2 \to N$ process, the beam functions $B_{a,b}$ contain the parton distributions
and perturbative collinear initial-state radiation from the colliding hard
partons, and the jet functions $J_i$ describe energetic collinear final-state
radiation from the primary $N$ hard partons produced in the collision. The soft
function $S_N$ describes the soft radiation in the event that couples to the in-
and outgoing hard partons. Since the collinear and soft radiation are not
separately observable, the soft function is convolved with the beam and jet
functions. The veto on additional jets restricts the collinear initial-state radiation, the final-state radiation, and
the soft radiation, which means the precise definition of the required beam,
jet, and soft functions depends on the veto variable.

For the case of an exclusive $0$-jet cross section, inclusive beam
functions can be obtained by using a simple event-shape variable called beam
thrust~\cite{Stewart:2009yx} to veto central jets. This $0$-jet cross section
has been studied for Drell-Yan production in Ref.~\cite{Stewart:2010pd} and for
Higgs production in Ref.~\cite{Berger:2010xi}.  The generalization of beam thrust to processes with $N$ jets is $N$-jettiness,
$\Tau_N$, introduced in Ref.~\cite{Stewart:2010tn}. It is defined as
%%%
\begin{equation}\label{eq:TauN_def}
\Tau_N = \sum_k
\min_i \Bigl\{ \frac{2 q_i\cdot p_k}{Q_i} \Bigr\}
\,,\end{equation}
%%%
where $i$ runs over $a, b$ for the two beams and $1, \ldots, N$ for the
final-state jets, the $q_i$ are massless reference momenta
for the jets and beams, and the $Q_i$ are normalization factors.
For the final-state jets we
can take $q_i^\mu = E_i\, (1,\vec n_i)$ where $E_i$ is the jet energy, and $\vec n_i$ is the jet direction.
For the initial-state jets we take $q_{a,b}^\mu = x_{a,b} \Ecm/2\, (1,\vec n_{a,b})$, where $x_{a,b}$ are the momentum fractions of the colliding partons, $\Ecm$ is the collider center-of-mass energy and $\vec n_{a,b}$ the beam directions.
It is convenient to define the dimensionless reference momenta and
their invariant products $ \hq_i^\mu = q_i^\mu /Q_i  \,, \ \hs_{ij} = 2\hq_i\cdot\hq_j$.

$N$-jettiness is designed such that in
the limit $\Tau_N \to 0$ the final state consists of $N$ narrow jets plus two
narrow initial-state radiation jets along the beam axis.  We can veto events with more than N jets by imposing a cut $\Tau_N < \Tau_N^{\rm cut} \ll Q_i$. The use of an event
shape for the jet veto makes a resummation of large logarithms to next-to-next-to-leading logarithmic (NNLL) order possible.
%%%
\begin{figure*}[t!]
\includegraphics[width=0.33\textwidth]{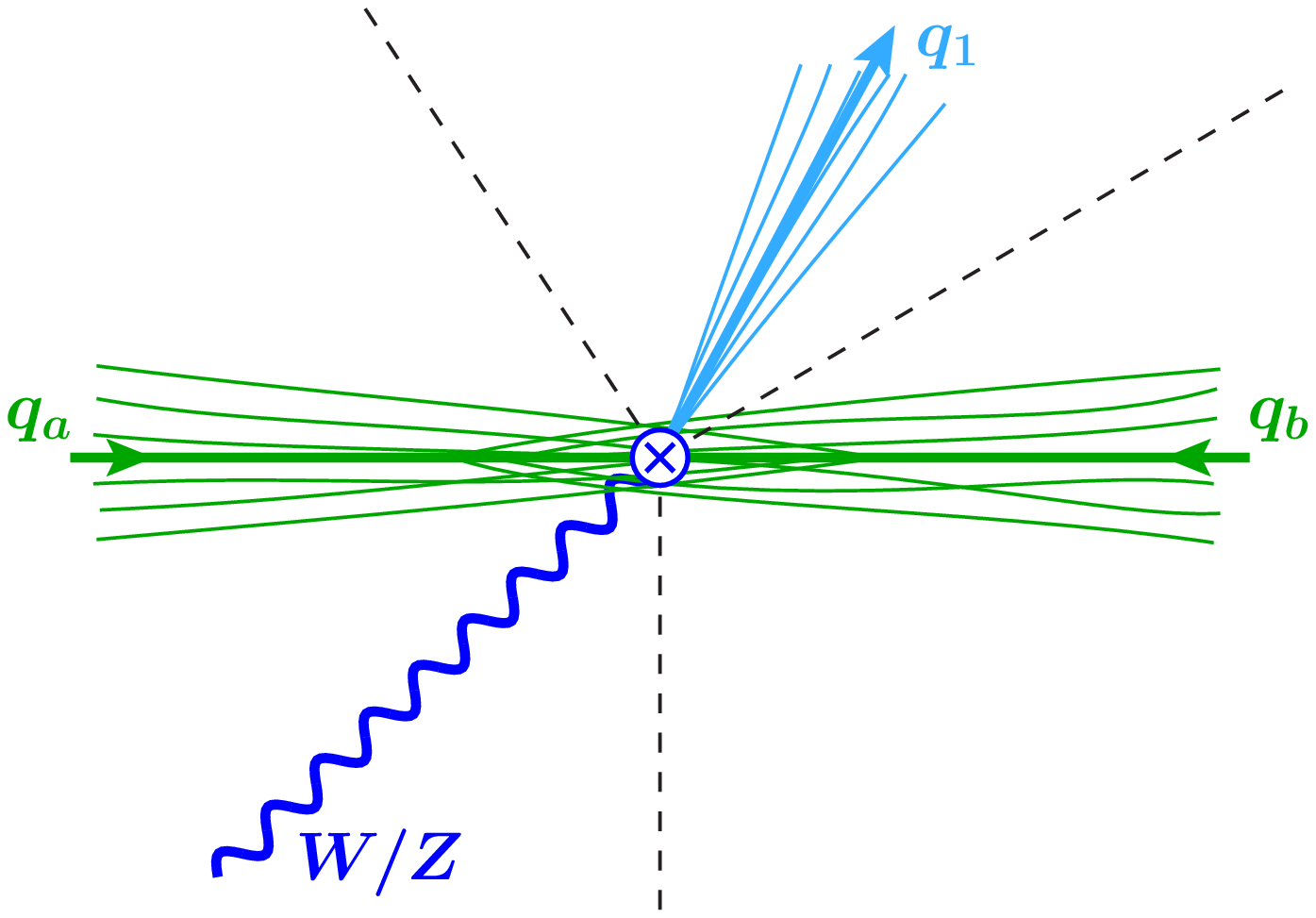}%
\hfill%
\includegraphics[width=0.33\textwidth]{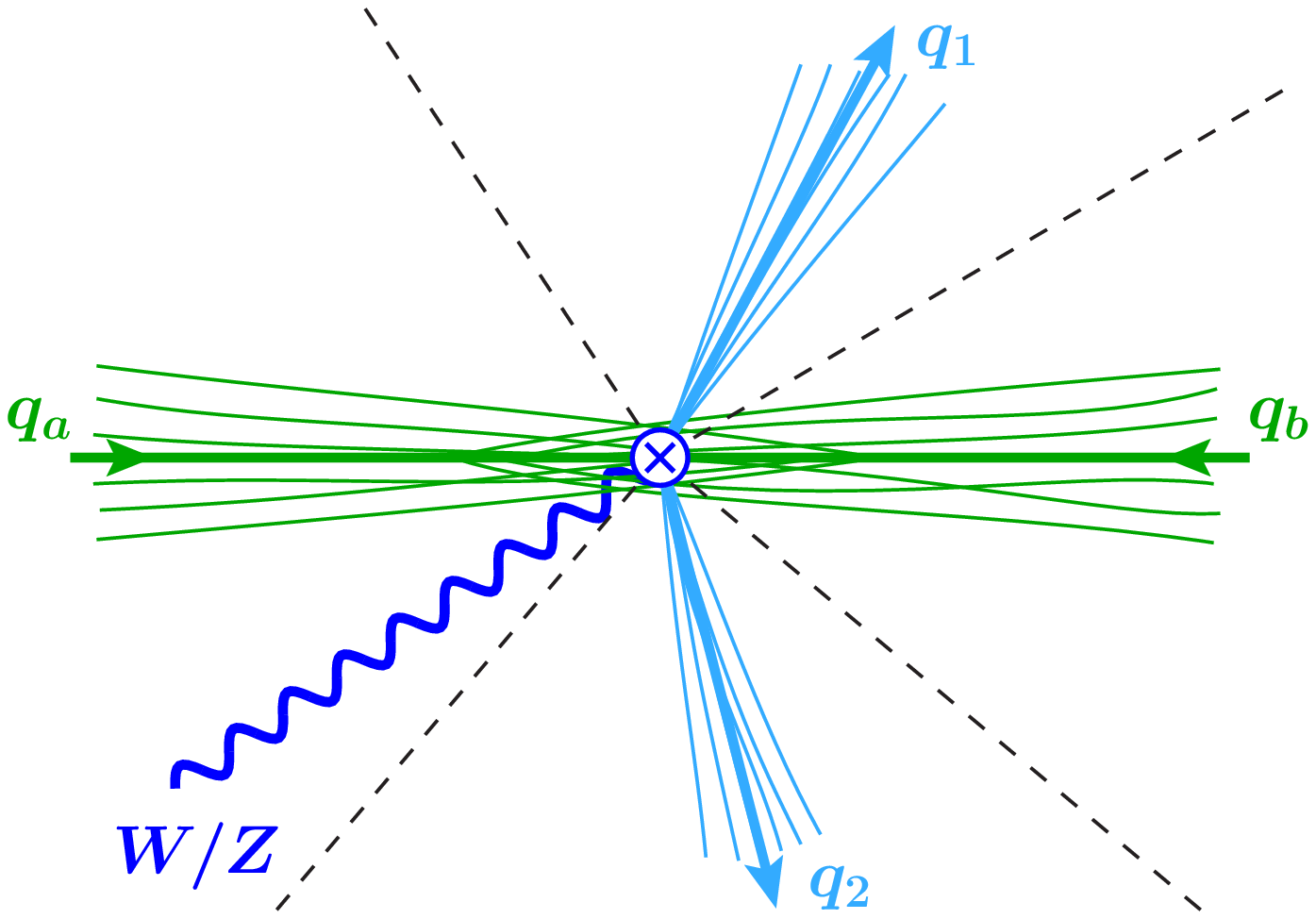}%
\hfill%
\includegraphics[width=0.33\textwidth]{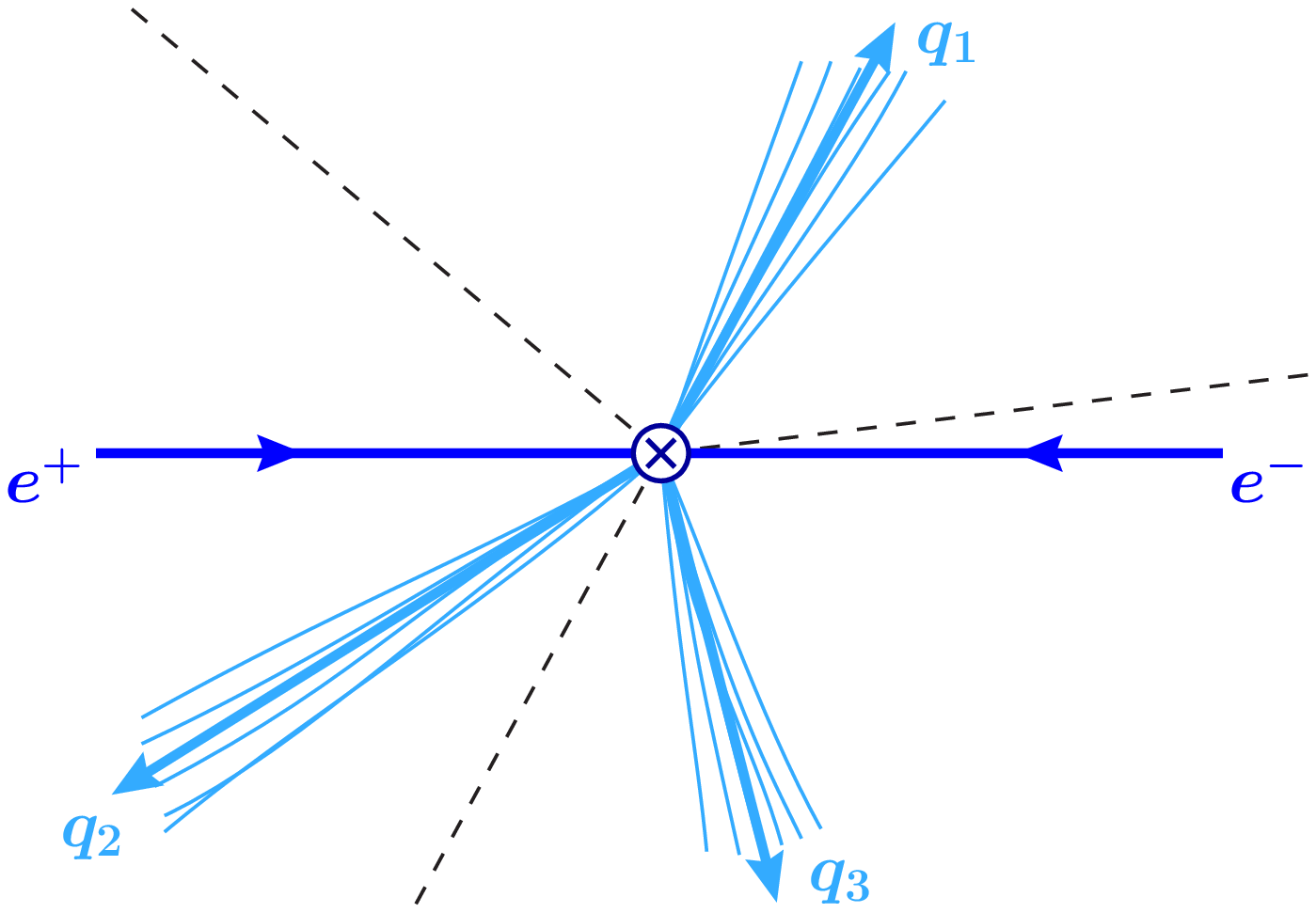}%
\vspace{-0.5ex}
\caption{ Jet and beam reference momenta for $1$-jettiness (left), $2$-jettiness (middle) and $e^+e^-$ $3$-jettiness (right). In the middle plot the jets and beams do not necessarily lie in a plane.\vspace{-1ex}}
\label{fig:jettiness}
\end{figure*}
\begin{figure*}[t!]
\hfill%
\includegraphics[width=0.33\textwidth]{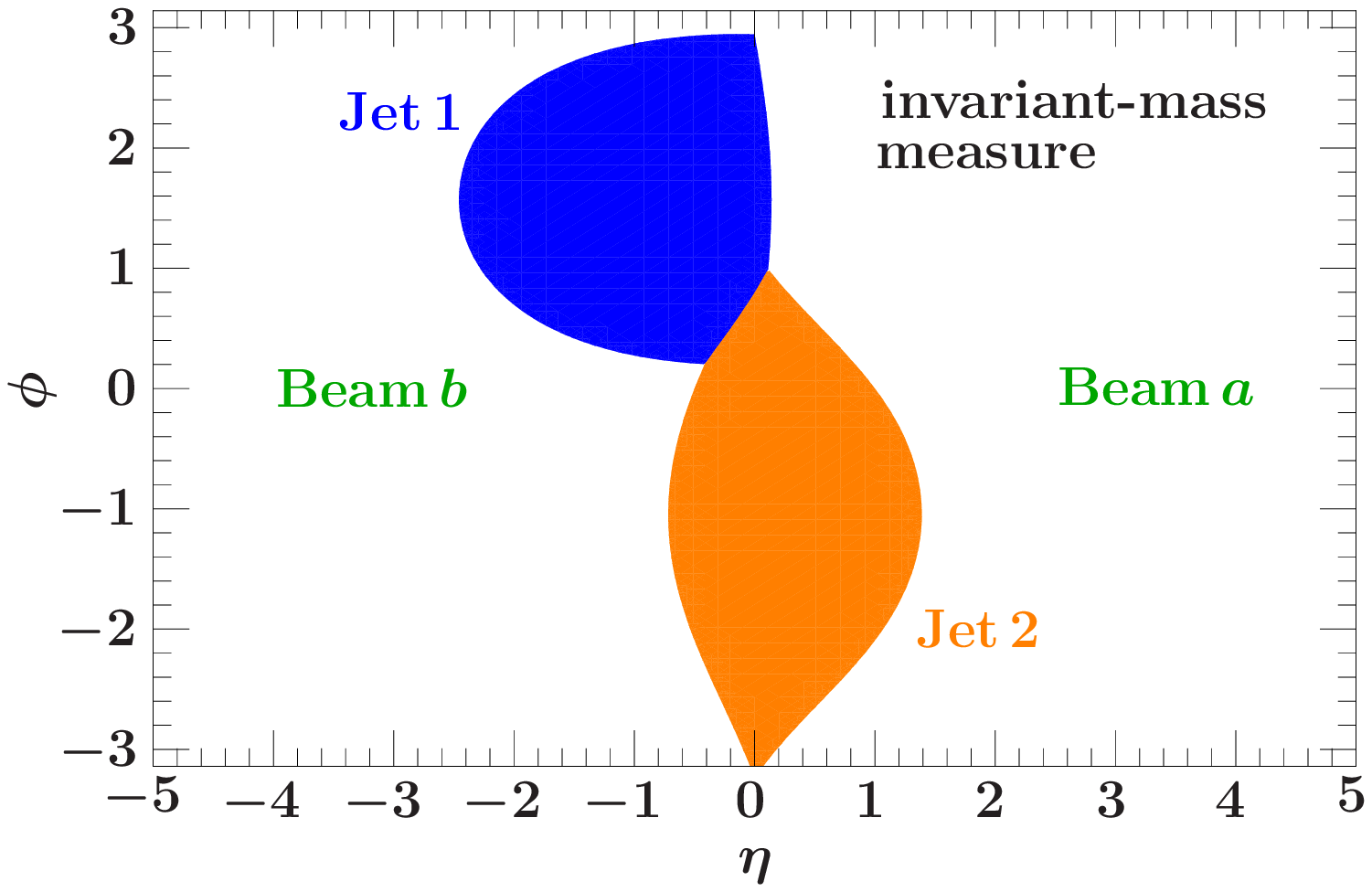}%
\hfill%
\includegraphics[width=0.33\textwidth]{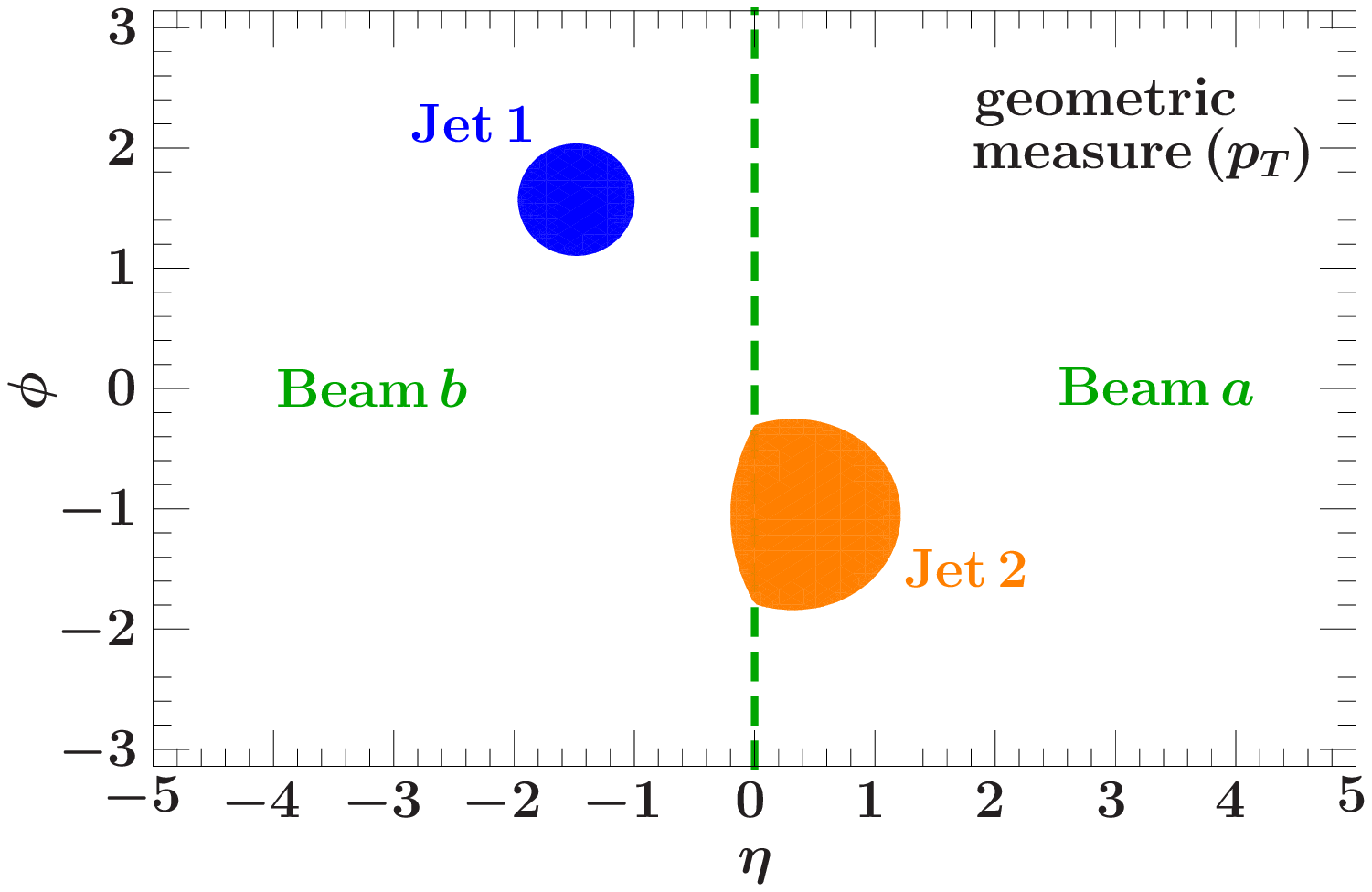}%
\hfill%
\includegraphics[width=0.33\textwidth]{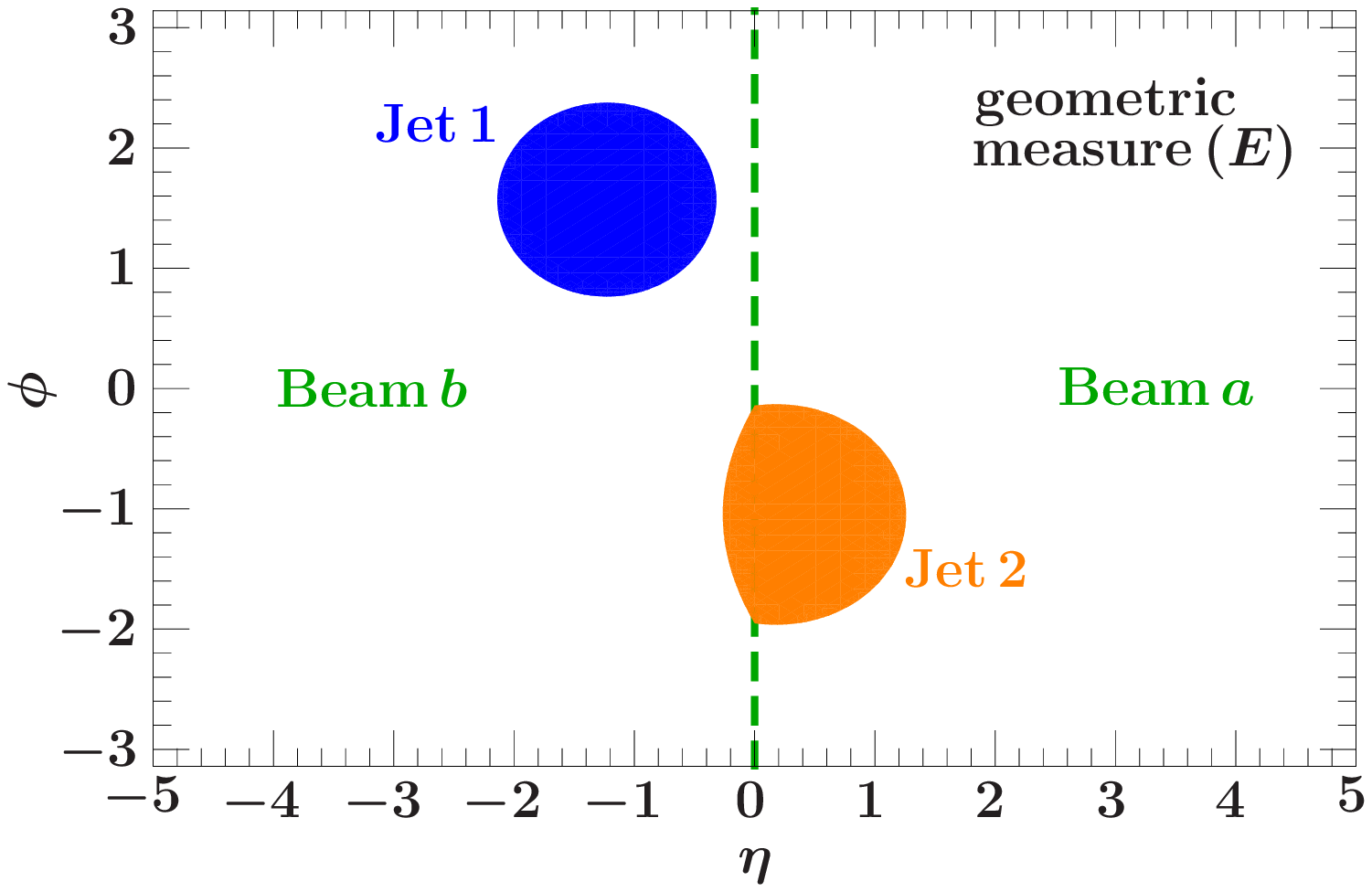}%
\hspace*{\fill}%
\vspace{-0.5ex}
\caption{Jet and beam regions for two jets using $2$-jettiness. On the left we use the invariant-mass measure $Q_i = Q$. In the middle we use the geometric measure $(p_T)$ with  $Q_i = | \vec{q}_{iT}| $ for the jets and $Q_{a,b} = x_{a,b}\Ecm$ for the beams. On the right we use the geometric measure $(E)$ with  $Q_i = E_i $ for the jets and $Q_{a,b} = x_{a,b}\Ecm$ for the beams. The first two cases were considered in Ref.~\cite{Jouttenus:2011wh}.\vspace{-3ex}}
\label{fig:2jet_etaphi}
\end{figure*}
%%%
$N$-jettiness assigns all particles to one of $N+2$ regions,
corresponding to the $N$ jets and $2$ beams, as seen in \fig{jettiness}.
Thus, $\Tau_N$ acts like a jet algorithm and different normalization
factors $Q_i$ result in different shapes for the jet regions, as discussed in
Ref.~\cite{Jouttenus:2011wh}. In addition the jet regions depend only on the reference
vectors $q_i^\mu$, but not on the distribution of soft radiation.
Three interesting choices are shown in \fig{2jet_etaphi}. The invariant-mass
measure is defined by $Q_i = Q$, where $Q^2$ is the total invariant mass-squared
of the hard interaction. Geometric measure $(p_T)$ is given by $Q_i =
| \vec{q}_{iT}| $ for the jets and $Q_{a,b} = x_{a,b}\Ecm$ for the beams, where
$| \vec{q}_{iT}|$ is the transverse momentum of jet $i$. Geometric measure
$(E)$ is like the $(p_T)$ case except that $Q_i = E_i$ for the jets. This ensures
that the jet regions remain of comparable size for jets at different rapidities.
For the geometric measure, the $x_{a,b}$ dependence cancels between $Q_{a,b}$ and
$q_{a,b}$, which is useful for the case of missing energy when the $x_{a,b}$ cannot be measured.

We can consider distinct measurements on each of these
``jets''. The simplest example is $\Tau_N^i$, the $N$-jettiness contribution
from each region $i$, where $\Tau_N = \sum_i \Tau_N^i$.  If the jet direction is chosen to be aligned
with the direction of the jet three-momentum,
then $N$-jettiness measures the mass of the jet \cite{Jouttenus:2011wh},
%%%
\begin{equation}
M_i^2 = P_i^2 = Q_i \Tau_N^i
\, .
 \end{equation}
%%%
Such an alignment can be achieved by minimizing geometric $N$-jettiness to find the jet direction \cite{Thaler:2011gf}.

Our soft-function calculation provides the last missing ingredient to obtain the exclusive $N$-jet \pagebreak[1] cross section resummed to NNLL for any process where the corresponding SCET hard function at NLO is known from the one-loop QCD calculation. The factorization theorem for the cross section combines all the ingredients together:
%%%
\begin{align} \label{eq:sigma_TauN}
\frac{\df\sigma}{\df \Tau_N^a\, \df\Tau_N^b\dotsb\df\Tau_N^N}
&=
\int\!\df x_a \df x_b \int\!\df^4 q\,\df\Phi_L(q) \int\! \df \Phi_N(\{q_J\})\,
M_N(\Phi_N, \Phi_L)\, (2\pi)^4 \delta^4\bigl(q_a + q_b - q_1 - \dotsb - q_N - q\bigr)
\nn\\ &\quad \times
\sum_{\kappa}
\int\!\df t_a\, B_{\kappa_a}(t_a, x_a, \mu)
\int\!\df t_b\, B_{\kappa_b}(t_b, x_b, \mu)
\prod_{J=1}^N \int\!\df s_J\, J_{\kappa_J}(s_J, \mu)
\\\nn &\quad \times
\vC_N^{\kappa\dagger}(\Phi_N, \Phi_L, \mu)\,
\hS_N^{\ \kappa} \biggl(\Tau_N^a - \frac{t_a}{Q_a}, \Tau_N^b - \frac{t_b}{Q_b}, \Tau_N^1 - \frac{s_1}{Q_1}, \ldots, \Tau_N^N - \frac{s_N}{Q_N} , \{\hq_i\}, \mu\biggr)\,
\vC_N^{\kappa}(\Phi_N, \Phi_L, \mu)
\, ,\end{align}
%%%
where the details of the notation, the running of the different components and the steps of the new calculation for the soft function are given in Ref.~\cite{Jouttenus:2011wh}. The factorization theorem for $N$-jettiness features inclusive jet and beam functions.

We employ a new calculational approach where we analytically extract the UV divergences by splitting the phase space into hemispheres depending on which Wilson lines the soft gluon attaches to. The so-called "hemisphere contributions" reproduce the anomalous dimension of the soft function as expected from the consistency of the factorization theorem. The remaining "non-hemisphere contributions", which encode the dependence on the boundaries between the regions, are reduced to one-dimensional numerical integrals.
The renormalized soft function can be written as
%%%
\begin{align} \label{eq:Sij_def}
\hS_N(\{k_i\}, \mu)
&= \id \prod_i \delta(k_i) + \sum_{i\neq j} \bT_i\cdot \bT_j\, \Bigl[
S_{ij,\hemi}^\one(\{k_i\}, \mu) + \sum_{m \neq i,j} S_{ij,m}^\one(\{k_i\}, \mu)
\Bigr]
+ \ord{\alpha_s^2}
\,,\end{align}
%%%
where $k_i$ are the soft contributions to $\Tau^i_N$. The final result for the renormalized hemisphere contribution is given by
%%%
\begin{align}
S_{ij,\hemi}^\one(\{k_i\}, \mu)
&= \frac{\alpha_s(\mu)}{4\pi}
\biggl[\frac{8}{\sqrt{\hs_{ij}}\, \mu}\cL_1\biggl(\frac{k_i}{\sqrt{\hs_{ij}}\,\mu}\biggr)
 - \frac{\pi^2}{6}\,\delta(k_i)\biggr]
 \prod_{m\neq i}\delta(k_m)
\,.\end{align}
%%%
The non-hemisphere contribution is given by
%%%
\begin{align} \label{eq:SNijm}
S_{ij,m}^\one(\{k_i\}, \mu)
&= \frac{\alpha_s(\mu)}{\pi}
\biggl\{
I_0\Bigl(\frac{\hs_{jm}}{\hs_{ij}}, \frac{\hs_{im}}{\hs_{ij}}, \Bigl\{\frac{\hs_{jl}}{\hs_{jm}}, \frac{\hs_{il}}{\hs_{im}}, \phi_{lm} \Bigr\}_{l\neq i,j,m} \Bigr)
\biggl[
\frac{1}{\mu}\cL_0\Bigl(\frac{k_i}{\mu}\Bigr)\, \delta(k_m) - \delta(k_i)\,\frac{1}{\mu}\cL_0\Bigl(\frac{k_m}{\mu}\Bigr)
\nn\\ & \quad
+ \ln\frac{\hs_{jm}}{\hs_{ij}} \,\delta(k_i)\, \delta(k_m) \biggr]
+ I_1\Bigl(\frac{\hs_{jm}}{\hs_{ij}}, \frac{\hs_{im}}{\hs_{ij}}, \Bigl\{\frac{\hs_{jl}}{\hs_{jm}}, \frac{\hs_{il}}{\hs_{im}}, \phi_{lm} \Bigr\}_{l\neq i,j,m} \Bigr)\,
\delta(k_i)\,\delta(k_m) \biggr\} \prod_{l\neq i,m} \delta(k_l)
\,,\end{align}
%%%
where $\phi_{lm}$ are the angles between the $\vec{\hq}_{l\perp}$ and
$\vec{\hq}_{m\perp}$ and the phase-space integrals $I_0, I_1$ are given in
Ref.~\cite{Jouttenus:2011wh}. An alternative approach to calculating $N$-jet
soft functions is discussed in Ref.~\cite{Bauer:2011hj}. $N$-jettiness has also
been extended to a jetshape, $N$-subjettiness, to identify boosted hadronic
objects~\cite{Thaler:2011gf,subjettiness}.

Our work in progress includes studying the phenomenology of $N$-jettiness to
predict jet mass spectra for exclusive $N$-jet events. One might want to study
the mass spectrum of the hardest jet, the average mass of all the jets or some
other combination of jet masses. These can be compared to experimental
measurements in order to improve our understanding of both the Standard Model
and any new physics that might be found at the LHC and Tevatron.

This work was supported in part by the Office of Nuclear Physics of the U.S. Department of Energy under the grant DE-FG02-94ER40818 and DE-FG02-90ER40546, and by the Department of Energy under the grant DE-SC003916. T.J. is also supported by a LHC-TI grant under the NSF grant PHY-0705682.

\bibliographystyle{aipproc}   % if natbib is available
%\bibliographystyle{aipprocl} % if natbib is missing

% \bibliography{../pp}

\end{document}